\documentclass[twocolumn,prb,amsmath,amssymb,floatfix,superscriptaddress]{revtex4-1}
\usepackage{color}
\usepackage{graphicx}
\usepackage{dcolumn}
\usepackage{bm}
\usepackage{times}
\begin{document}

\title{Neutron inelastic scattering measurements of low energy phonons in the multiferroic BiFeO$_3$}

\author{John~A.~Schneeloch}
\affiliation{Condensed Matter Physics and Materials Science Department, Brookhaven National Laboratory, Upton, New York 11973, USA}
\affiliation{Department of Physics and Astronomy, Stony Brook University, Stony Brook, New York 11794, USA}
\author{Zhijun~Xu}
\affiliation{Condensed Matter Physics and Materials Science Department, Brookhaven National Laboratory, Upton, New York 11973, USA}
\affiliation{Physics Department, University of California, Berkeley, CA 94720, USA}
\affiliation{Materials Science Division, Lawrence Berkeley National Laboratory, Berkeley, CA, 94720, USA}
\author{Jinsheng~Wen}
\affiliation{Physics Department, University of California, Berkeley, CA 94720, USA}
\affiliation{Materials Science Division, Lawrence Berkeley National Laboratory, Berkeley, CA, 94720, USA}
\author{P. M. Gehring}
\affiliation{NIST Center for Neutron Research, National Institute of
Standards and Technology, Gaithersburg, Maryland 20899, USA}
\author{C. Stock}
\affiliation{School of Physics and Astronomy, University of Edinburgh, Edinburgh EH9 3JZ, United Kingdom}
\author{M. Matsuda}
\author{B. Winn}
\affiliation{Quantum Condensed Matter Division, Oak Ridge National Laboratory,
Oak Ridge, TN, 37831, USA}
\author{Genda~Gu}
\author{Stephen M. Shapiro}
\affiliation{Condensed Matter Physics and Materials Science
Department, Brookhaven National Laboratory, Upton, New York 11973,
USA}
\author{R. J. Birgeneau}
\affiliation{Physics Department, University of California, Berkeley, CA 94720, USA}
\affiliation{Materials Science Division, Lawrence Berkeley National
Laboratory, Berkeley, CA, 94720, USA}
\author{T. Ushiyama}
\author{Y. Yanagisawa}
\author{Y. Tomioka}
\author{T. Ito}
\affiliation{National Institute of Advanced Industrial Science and Technology (AIST), Tsukuba, Ibaraki 305-8562, Japan}
\author{Guangyong~Xu}
\affiliation{Condensed Matter Physics and Materials Science
Department, Brookhaven National Laboratory, Upton, New York 11973,
USA}
\date{\today}

\begin{abstract}

We present neutron inelastic scattering measurements of the low-energy phonons in single crystal BiFeO$_3$. The dispersions of the three acoustic phonon modes (LA along [100], TA$_1$ along [010] and TA$_2$ along $[1\bar{1}0]$) and two low energy optic phonon modes (LO and TO$_1$) have been mapped out between 300\,K and 700\,K.  Elastic constants are extracted from the phonon measurements.  The energy linewidths of both TA phonons at the zone boundary clearly broaden when the system is warmed toward the magnetic ordering temperature $T_N = 640$\,K. This suggests that the magnetic and low-energy lattice dynamics in this multiferroic material are coupled.
\end{abstract}

\maketitle

\section{Introduction}

Being both ferroelectric and antiferromagnetic at room temperature, BiFeO$_3$ holds tantalizing promise for novel device applications and has become one of the most studied multiferroic materials.~\cite{Catalan}  The direct coupling between the static ferroelectric order ($T_C \approx 1100$\,K)~\cite{BFO_Structure1,BFO_Structure2} and antiferromagnetic order ($T_N \approx 640$\,K) is relatively weak, resulting in a cycloidal spin structure with long wavelength modulations ($\lambda \approx 620$~\AA).~\cite{spiral} Yet there has been a growing interest in the coupling between dynamical properties in recent years.  For example, a ``magneto-optical'' resonance has been proposed theoretically,~\cite{sousa:012406} and extra modes located below the lowest-lying optic phonon energies have been observed in Raman measurements.~\cite{Cazayous,Kumar1,Rovillain,BiFeO3_film_field}  Zone-center optic phonon modes have been studied extensively using Raman spectroscopy~\cite{BFO_Raman1,BFO_Raman2,haumont:132101,Rovillain_Phonon} and infrared reflectivity.~\cite{Lobo_infrared}  A full understanding of the energy-momentum dependence of the dynamics in BiFeO$_3$ is, however, still far from complete.  Only with the recent achievements in the growth of large BiFeO$_3$ single crystals have researchers been able to start carrying out neutron inelastic and x-ray scattering measurements on both the lattice and spin dynamics~\cite{BFO_Xu,cheong,Matsuda,Delaire,BFO_IXS}.

In this paper we present neutron inelastic scattering measurements of the low-energy phonon modes in a large BiFeO$_3$ single crystal from 300\,K to 700\,K.  The dispersions of the three acoustic phonon modes (TA$_1$, propagating along [010] and polarized along [100]; TA$_2$, propagating along $[1\bar{1}0]$ and polarized along [110]; and LA, propagating and polarized along [100]), and the lowest-lying optic modes are mapped out in three different Brillouin zones. The measured phonon dispersions and intensities (after removing the Bose-factor) show no significant change with temperature outside of experimental errors.  However, the transverse acoustic modes at the zone-boundary clearly broaden in energy when the sample is heated towards $T_N$; this suggests that a coupling exists between the low-energy lattice and antiferromagnetic dynamics in this multiferroic system.

\section{Experiment}

Single crystals of BiFeO$_3$ were grown using the traveling-solvent floating-zone technique.~\cite{Ito_BFO}  The crystal used in this study has a cylindrical shape, a mass of 4\,g, and a mosaic 
full width at half-maximum (FWHM) of less than 1$^\circ$.  Neutron inelastic scattering experiments were performed on the HYSPEC time-of-flight spectrometer~\cite{HYSPEC}, which is located at the Spallation Neutron Source (SNS) at Oak Ridge National Laboratory.  The incident neutron energy $E_i$ was fixed at 20\,meV.  We will refer to the pseudo-cubic unit cell with $a=3.9$\,\AA\ when describing our results.  In this coordinate system the magnetic ordering wave vector is (0.5,0.5,0.5).  The crystal $c$-axis was oriented vertically to allow access to reflections of the form $(HK0)$ in the horizontal scattering plane.  During the measurements the crystal was rotated about the $c$-axis by 90$^\circ$ in steps of 2$^\circ$ in order to provide full coverage of the reciprocal lattice scattering plane (within kinematical constraints).

\section{Results and Discussion}

\begin{figure}[ht]
\includegraphics[width=\linewidth,trim=0cm -1cm 0cm -1cm,clip]{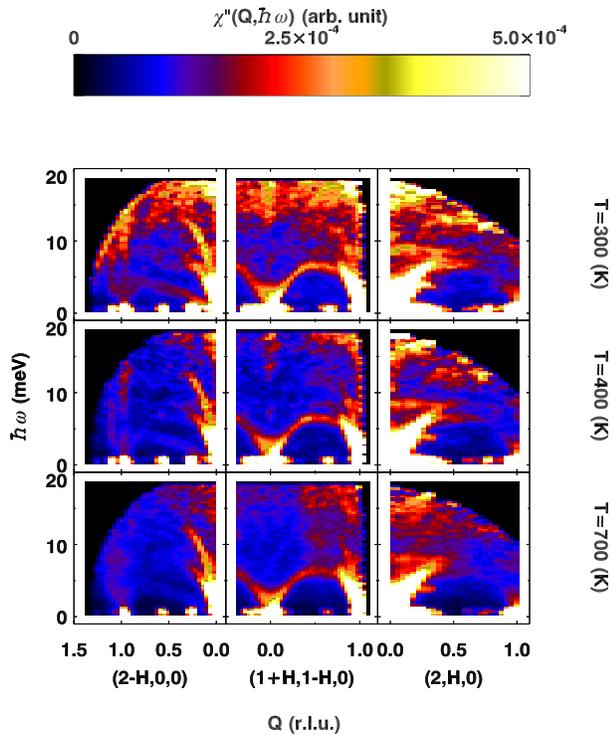}
\caption{(Color online) The dynamic response function $\chi''(Q,\hbar\omega)$ of single crystal BiFeO$_3$, at 300\,K (top row), 500\,K (middle row), and 700\,K (bottom row).  The intensities are shown in the $(HK0)$ plane,
for $q$ along [100] (left column), $[1\bar{1}0]$ (middle column), and [010] (right column), respectively.  The strong intensities visible near the top of each panel are spurious and occur when $\hbar\omega$ approaches $E_i$. These spurious intensities are temperature independent.  They appear stronger at 300\,K (top row) because the data have been divided by the Bose factor.} \label{fig:1}
\end{figure}

In Fig.~\ref{fig:1}, we plot the dynamic response function $\chi''({\bf Q},\hbar\omega)= (1-e^{-\hbar\omega/k_BT})S({\bf Q},\hbar\omega)$, which is simply the measured neutron scattering intensity divided by the Bose factor, at 300, 500, and 700\,K.  Each panel contains data obtained from two-dimensional (2D) slices through energy-momentum space. The left panels show LA phonons propagating along [100] between the $\Gamma$ points at (100) and (200). The center panels show phonons propagating along $[1\bar{1}0]$ between (110) and (200).  Near the (110) zone center these are TA$_2$ and TO$_2$ modes. The right panels show phonons propagating along [010] between (200) and (210), which are TA$_1$ and TO$_1$ phonons near (200).  We see that whereas the LA, TA$_1$, and TA$_2$ phonons are all well defined, the optic modes are much less clear in our measurements.  In particular, the structure factor of the TO$_2$ mode near (110) is extremely weak and can barely be observed.

The data in Fig.~\ref{fig:1} reveal no significant changes in the overall features of the dynamic response in BiFeO$_3$ between 300\,K and 700\,K.  To obtain more detailed information, we fit these data using Lorentzian functions of $q$ convoluted with the Gaussian instrumental resolution function to extract the acoustic and optic phonon intensities vs. energy transfer $\hbar\omega$ at various wave vectors $q$. The phonon energy, energy width ($2\Gamma$), and intensity were also obtained from these fits.

\begin{figure}[ht]
\includegraphics[width=0.9\linewidth,angle=90,trim=-2cm -2cm 0cm 0cm,clip]{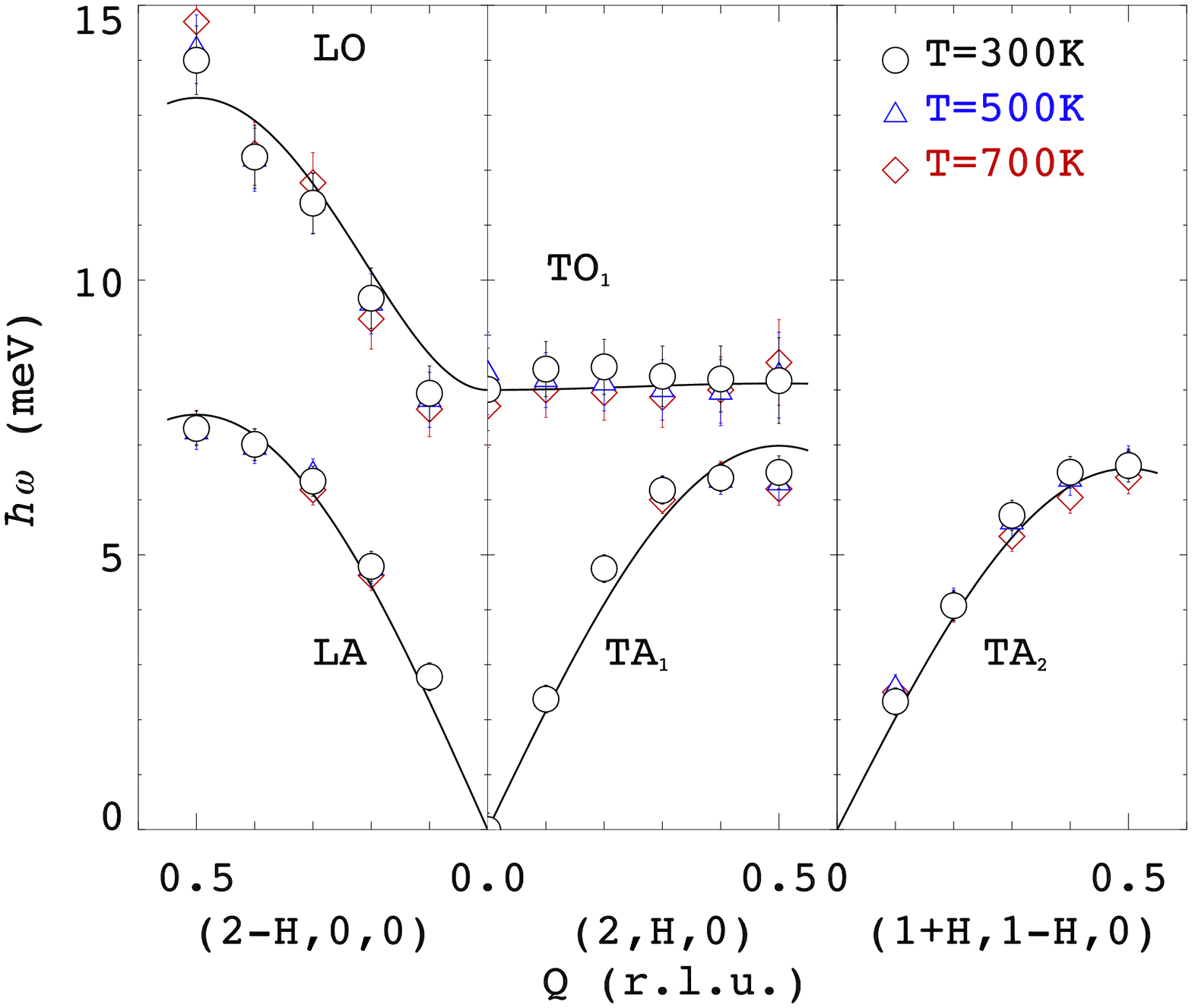}
\caption{(Color online) Phonon dispersions measured along various symmetry directions. The phonon energies are obtained from the fits described in the text.  Lines are guides to the eye.} \label{fig:2}
\end{figure}

The phonon dispersions obtained from the fits are shown in Fig.~\ref{fig:2}.  We note here that all uncertainties and error bars in this and subsequent figures represent standard deviations.  The energies of all acoustic modes show little to no temperature variation, especially in the long wave-length (small $q$) limit.  Near the zone boundary a slight softening on warming is observed for the TA$_1$ mode near (2,0.5,0) and for the TA$_2$ mode near (1.5,0.5,0).  However the size of this softening is $\alt 0.3$\,meV, which is comparable to our experimental uncertainty.  The optic phonon intensities are considerably weaker than those of the acoustic phonons, and consequently the optic phonon energies have larger error bars.  The TO$_2$ phonon intensity near the (110) zone center is particularly weak (Fig.~\ref{fig:1}, middle column), and for this reason we were not able to determine the TO$_2$ phonon dispersion along $[1\bar{1}0]$.  In the other two zones, the TO$_1$ mode propagating along [100] is almost flat, while the LO mode measured along [100] is quite steep.  Similar to the negligible softening observed for the zone-boundary acoustic modes, 
the zone-center optic modes also show a slight decrease in energy on warming.  But the zone-boundary optic mode energies exhibit a slight increase instead.

Our results are generally consistent with the neutron inelastic scattering measurements of Delaire {\em et al}.\ on powder samples of BiFeO$_3$.~\cite{Delaire}  They found that the lowest-energy optic phonon branch bottoms out around 8\,meV and that the top of the TA branch is close to 6.5\,meV.  From Fig.~\ref{fig:2}, one can see that the relevant energies are roughly 8.0\,meV (lowest TO phonon energy) and 6.5\,meV (top of the TA phonon branch).  The value of 8.0\,meV is also close to the 57\,cm$^{-1}$ line observed via Raman scattering~\cite{BFO_Raman2} and the 66\,cm$^{-1}$ line seen in infrared measurements,~\cite{Lobo_infrared} both of which probe zone-center ($q=0$) optic modes.  Other Raman~\cite{BFO_Raman1} and x-ray inelastic scattering studies ~\cite{BFO_IXS} suggest that the lowest-energy TO mode is around 9.2\,meV.  However our results show that a zone-center optic mode is present at 8\,meV, clearly below 9\,meV, as is illustrated 
in Figs.~\ref{fig:1} and \ref{fig:2}.

\begin{figure}[ht]
\includegraphics[width=\linewidth,trim=-2cm -2cm 0cm -2cm,clip]{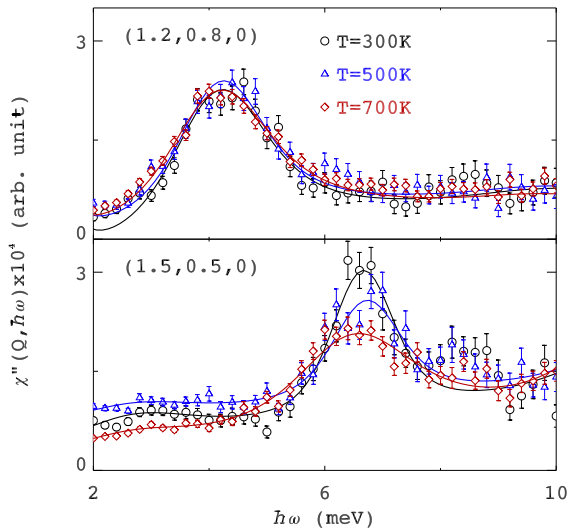}
\caption{(Color online) Energy cuts at $Q=(1.2,0.8,0)$ (top frame) and $Q=(1.5,0.5,0)$ (bottom frame).  Plotted intensities correspond to $\chi''(Q,\hbar\omega)$ measured at 300\,K (black circles), 500\,K (blue triangles), and 700\,K (red diamonds).} \label{fig:4}
\end{figure}

Based on the measured acoustic phonon branches, a number of useful bulk parameters such as the elastic constants can be obtained.~\cite{ElasticConstant}  Here we make use of the expression
\begin{equation}
v_i=\sqrt{\frac{C_{eff}}{\rho}},
\end{equation}
where $v_i$ is the phonon velocity for the TA$_1$, TA$_2$ or LA modes, $\rho=8.408$\,g/cm$^3$ is the
density of BiFeO$_3$, and $C_{eff}$ is the effective elastic constant, where $C_{eff}=C_{11}$ for the LA mode along [100], $C_{44}$ for the TA$_1$ mode along [100], and $(C_{11}-C_{12})/2$ for the TA$_2$ mode along $[1\bar{1}0]$.

\begin{table}[ht]
\caption{Description of the acoustic phonons.}
\begin{ruledtabular}
\begin{tabular}{cccc}
 Mode & Propagation vector  & Polarization vector & Velocity (m/s)\\
\hline
LA& [100] & [100] &2.6(5)$\times 10^3$\\
TA$_1$ &[100]& [010] & 2.2(4)$\times 10^3$\\
TA$_2$ &$[1\bar{1}0]$ & [110] & 1.6(3)$\times 10^3$\\
\end{tabular}
\end{ruledtabular}
\label{tab:1}
\end{table}

The acoustic phonon velocities were obtained from the limiting slopes of the 300\,K dispersion curves shown in Fig.~\ref{fig:2} and are are listed in Table~I.  From these we obtain $C_{11}=58(6)$\,GPa, $C_{44}=42(4)$\,GPa, $C_{12}=17(2)$\,GPa, and the bulk modulus $B=\frac{1}{3}(C_{11}+2C_{12})=31(3)$\,GPa.  We note that the room temperature values of the longitudinal elastic constant $C_{11}$ derived from previous experimental~\cite{BFO_IXS,Smirnova_Acoustic,DoigK_phonon} and theoretical work~\cite{Goffinet,Shang_BFOelastic} on BiFeO$_3$ are highly inconsistent.  The value of $C_{11}$ ranges from $\approx 60$\,GPa (our neutron measurements), to $\approx 125$\,GPa (ultrasonic measurements), to $\approx 207$\,GPa (x-ray inelastic measurements). We believe that the difference between our results and those from the ultrasonic measurements is an artifact of comparing $C_{11}$ for a single crystal sample to $C_L$ for a ceramic sample.  $C_{11}$ is determined by the longitudinal acoustic phonon velocity along [100].  When the LA phonon velocities along other crystallographic directions are higher, a powder/ceramic average will lead to a larger value of $C_L$.  With 
respect to the x-ray scattering measurements in Ref.~\onlinecite{BFO_IXS}, we note that a different coordinate system with trigonal symmetry was used 
along with a different definition of $C_{11}=\rho v_{LA[110]}^2$. For the purpose of obtaining bulk properties in the pseudo-cubic coordinate system, our results provide a direct reference for the acoustic phonon velocities and elastic constants.

In addition to the acoustic modes and low-energy optic modes discussed above, we have also observed a few spurious signals in our measurements.  For example, one can see a faint streak of intensity originating from the (200) Bragg peak in the top-left panel of Fig.~\ref{fig:1} that increases to about 4\,meV at (1.5,0,0), thus resembling a phonon branch below the LA branch.  The intensity of this spurion has no temperature dependence, which is why when $\chi''$ is plotted in Fig.~\ref{fig:1} (and the Bose factor is divided out) it appears to weaken at higher temperatures.  When we checked the energy gain side of our scattering data, we found no trace of this spurion, which does not happen with real phonons.  We believe it is most likely caused by spurious intensities coming from detectors located close to the strong (200) Bragg peak.  We also looked for, but did not detect, any extra modes that resemble the ``electromagnon'' excitations observed by Raman measurements~\cite{Cazayous,Kumar1,Rovillain,BiFeO3_film_field}.

\begin{figure}[ht]
\includegraphics[width=\linewidth,trim=-2cm -2cm 0cm -2cm,clip]{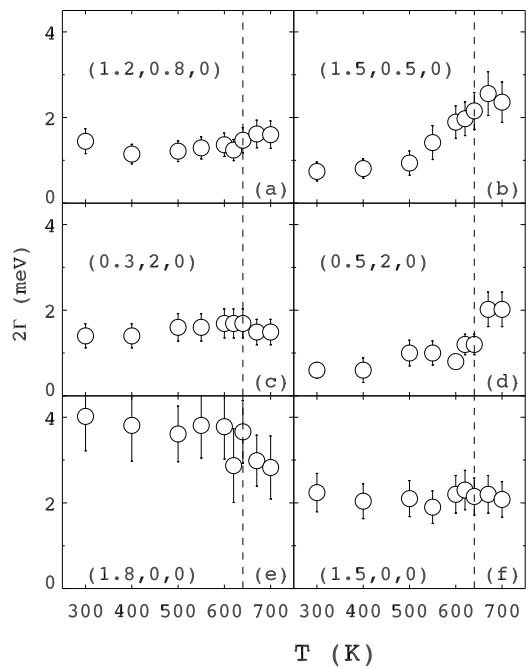}
\caption{(Color online) Energy widths (2$\Gamma$) of acoustic phonons measured at different $Q$. The dashed line in 
the figure denotes the N\`{e}el temperature $T_N$.}
\label{fig:3}
\end{figure}

Fig.~\ref{fig:4} shows energy scans for the TA$_2$ mode ($q \parallel [1\bar{1}0]$) at two different wave vectors. The intensities correspond to the dynamic response function $\chi``(Q,\omega)$ (measured scattering intensity divided by the Bose factor).  For small values of $q$, like that shown in the top panel, there is virtually no change of $\chi``(Q,\omega)$ between 300\,K and 700\,K.  But at the zone boundary ($q=0.5$), as shown in the bottom panel, we observe a clear broadening of the phonon linewidth that is accompanied with a slight softening as the temperature increases towards $T_N \approx 640$\,K.  We see some evidence of an optic phonon located near 8.5\,meV in the bottom panel of Fig.~\ref{fig:4}, however it is so weak that we are unable to fit it in a meaningful way.  Fortunately it does not affect our measurement of the TA$_2$ phonon linewidth.  We also observe a broadening of the zone-boundary TA$_1$ mode, but none is seen for the zone-boundary LA mode.  A plot of the phonon energy widths $2\Gamma$ between 300\,K and 700\,K for different $Q$ is shown in Fig.~\ref{fig:3}.  Here we see that only in panels (b) and (d), which are the zone boundary positions for the TA$_1$ and TA$_2$ modes, do the phonon energy widths increase on heating close to $T_N$.  For all other wave vectors there is very little temperature dependence to the phonon energy widths.

This anomaly in the TA phonon energy width is unlikely to be the result of pure thermal broadening because it correlates well with $T_N$ and is not seen at other wave vectors.  Previous neutron scattering measurements of the phonon density-of-states (DOS) in powdered BiFeO$_3$~\cite{Delaire} reveal a strong anharmonicity as evidenced by phonon softening and broadening on warming.  These 
DOS measurements are mostly focused on the optic modes; by contrast, our measurements show that the acoustic phonons in BiFeO$_3$ are well defined and display no 
significant energy broadening or softening from 300\,K to 700\,K for small $q$.  Signs of anharmonicity only occur for TA phonons at the zone-boundaries.

The fact that the TA phonon broadening at the zone boundary becomes pronounced when the N\`{e}el temperature $T_N$ is approached and the magnetic order starts to melt implies a possible connection between the TA phonons and the magnetic phase transition.  Previous Raman measurements~\cite{haumont:132101,Rovillain_Phonon} have suggested that certain optic phonon modes also behave anomalously near $T_N$, which is indicative of a spin-phonon coupling in this multiferroic material and possibly other multiferroic perovskites.~\cite{Stock_PFN}  Our results show that the effects of spin-phonon coupling also extend to the TA phonons. It is possible that this anomalous zone-boundary TA phonon broadening is a combination of spin-polar phonon (TO mode) coupling and TA-TO phonon coupling.  While the lowest-lying TO phonon can be affected by the magnetic order near $T_N$ based on results from the optic measurements, the coupling between the TA and TO modes can lead to the anomalies in the TA branches as well. The coupling is strongest at zone-boundaries because the energy difference between the two modes decreases at large $q$.  This can also explain the lack of energy broadening in the LA mode, as the LO phonon energy is much higher at zone boundary and the LO-LA coupling there should be significantly weaker.  Although the underlying mechanism of this coupling in BiFeO$_3$ is not entirely clear, it is evident that the melting of the magnetic order at $T_N$ correlates with a decrease in the zone-boundary TA phonon lifetime, which suggests the development of a lattice instability.

\section{Summary}
Using neutron time-of-flight spectroscopy we have measured the dispersions of the LA, TA$_1$, and TA$_2$ phonon modes for a single crystal of BiFeO$_3$ and determined the elastic constants $C_{11}$, $C_{12}$, and $C_{44}$.  The lowest energy TO and LO phonon dispersions were also measured, and these are consistent with previous Raman and infrared studies.  Our results suggest that 
the zone-center optic mode energy softens slightly on warming.  Most importantly, the transverse acoustic phonons measured at the zone boundary clearly broaden in energy when $T_N$ is 
approached from below.  This indicates that the acoustic phonon modes and magnetic excitations are coupled in this multiferroic material.

\section*{Acknowledgments}
JAS, ZJX, GDG, SMS, and GYX acknowledge support by Office of Basic Energy Sciences, U.S. Department of Energy under contract No. DE-AC02-98CH10886. JW and RJB are also supported by the Office of Basic Energy Sciences, U.S. Department of Energy through contract No. DE-AC02-05CH11231. This research at the Oak Ridge National Laboratory Spallation Neutron Source was sponsored by the Scientific User Facilities Division, Office of Basic Energy Sciences, U.\ S.\ Department of Energy. CS acknowledges the Carnegie Trust for the Universities of Scotland and the Royal Society.  TI is partly supported by the Mitsubishi Foundation.


\begin{thebibliography}{27}%
\makeatletter
\providecommand \@ifxundefined [1]{%
 \@ifx{#1\undefined}
}%
\providecommand \@ifnum [1]{%
 \ifnum #1\expandafter \@firstoftwo
 \else \expandafter \@secondoftwo
 \fi
}%
\providecommand \@ifx [1]{%
 \ifx #1\expandafter \@firstoftwo
 \else \expandafter \@secondoftwo
 \fi
}%
\providecommand \natexlab [1]{#1}%
\providecommand \enquote  [1]{``#1''}%
\providecommand \bibnamefont  [1]{#1}%
\providecommand \bibfnamefont [1]{#1}%
\providecommand \citenamefont [1]{#1}%
\providecommand \href@noop [0]{\@secondoftwo}%
\providecommand \href [0]{\begingroup \@sanitize@url \@href}%
\providecommand \@href[1]{\@@startlink{#1}\@@href}%
\providecommand \@@href[1]{\endgroup#1\@@endlink}%
\providecommand \@sanitize@url [0]{\catcode `\\12\catcode `\$12\catcode
  `\&12\catcode `\#12\catcode `\^12\catcode `\_12\catcode `\%12\relax}%
\providecommand \@@startlink[1]{}%
\providecommand \@@endlink[0]{}%
\providecommand \url  [0]{\begingroup\@sanitize@url \@url }%
\providecommand \@url [1]{\endgroup\@href {#1}{\urlprefix }}%
\providecommand \urlprefix  [0]{URL }%
\providecommand \Eprint [0]{\href }%
\providecommand \doibase [0]{http://dx.doi.org/}%
\providecommand \selectlanguage [0]{\@gobble}%
\providecommand \bibinfo  [0]{\@secondoftwo}%
\providecommand \bibfield  [0]{\@secondoftwo}%
\providecommand \translation [1]{[#1]}%
\providecommand \BibitemOpen [0]{}%
\providecommand \bibitemStop [0]{}%
\providecommand \bibitemNoStop [0]{.\EOS\space}%
\providecommand \EOS [0]{\spacefactor3000\relax}%
\providecommand \BibitemShut  [1]{\csname bibitem#1\endcsname}%
\let\auto@bib@innerbib\@empty
\bibitem [{\citenamefont {Catalan}\ and\ \citenamefont
  {Scott}(2009)}]{Catalan}%
  \BibitemOpen
  \bibfield  {author} {\bibinfo {author} {\bibfnamefont {G.}~\bibnamefont
  {Catalan}}\ and\ \bibinfo {author} {\bibfnamefont {J.~F.}\ \bibnamefont
  {Scott}},\ }\href@noop {} {\bibfield  {journal} {\bibinfo  {journal}
  {Advanced Materials}\ }\textbf {\bibinfo {volume} {21}},\ \bibinfo {pages}
  {2463} (\bibinfo {year} {2009})}\BibitemShut {NoStop}%
\bibitem [{\citenamefont {Palewicz}\ \emph {et~al.}(2007)\citenamefont
  {Palewicz}, \citenamefont {Przenioslo}, \citenamefont {Sosnowska},\ and\
  \citenamefont {Hewat}}]{BFO_Structure1}%
  \BibitemOpen
  \bibfield  {author} {\bibinfo {author} {\bibfnamefont {A.}~\bibnamefont
  {Palewicz}}, \bibinfo {author} {\bibfnamefont {R.}~\bibnamefont
  {Przenioslo}}, \bibinfo {author} {\bibfnamefont {I.}~\bibnamefont
  {Sosnowska}}, \ and\ \bibinfo {author} {\bibfnamefont {A.~W.}\ \bibnamefont
  {Hewat}},\ }\href@noop {} {\bibfield  {journal} {\bibinfo  {journal} {Acta
  Crystal. B}\ }\textbf {\bibinfo {volume} {63}},\ \bibinfo {pages} {537}
  (\bibinfo {year} {2007})}\BibitemShut {NoStop}%
\bibitem [{\citenamefont {Palewicz}\ \emph {et~al.}(2010)\citenamefont
  {Palewicz}, \citenamefont {Sosnowska}, \citenamefont {Przenioslo},\ and\
  \citenamefont {Hewat}}]{BFO_Structure2}%
  \BibitemOpen
  \bibfield  {author} {\bibinfo {author} {\bibfnamefont {A.}~\bibnamefont
  {Palewicz}}, \bibinfo {author} {\bibfnamefont {I.}~\bibnamefont {Sosnowska}},
  \bibinfo {author} {\bibfnamefont {R.}~\bibnamefont {Przenioslo}}, \ and\
  \bibinfo {author} {\bibfnamefont {A.~W.}\ \bibnamefont {Hewat}},\ }\href@noop
  {} {\bibfield  {journal} {\bibinfo  {journal} {Acta Physica Polonica A}\
  }\textbf {\bibinfo {volume} {117}},\ \bibinfo {pages} {296} (\bibinfo {year}
  {2010})}\BibitemShut {NoStop}%
\bibitem [{\citenamefont {Sosnowska}\ \emph {et~al.}(1982)\citenamefont
  {Sosnowska}, \citenamefont {Neumaier},\ and\ \citenamefont
  {Steichele}}]{spiral}%
  \BibitemOpen
  \bibfield  {author} {\bibinfo {author} {\bibfnamefont {I.}~\bibnamefont
  {Sosnowska}}, \bibinfo {author} {\bibfnamefont {T.~P.}\ \bibnamefont
  {Neumaier}}, \ and\ \bibinfo {author} {\bibfnamefont {E.}~\bibnamefont
  {Steichele}},\ }\href@noop {} {\bibfield  {journal} {\bibinfo  {journal}
  {Journal of Physics C: Solid State Physics}\ }\textbf {\bibinfo {volume}
  {15}},\ \bibinfo {pages} {4835} (\bibinfo {year} {1982})}\BibitemShut
  {NoStop}%
\bibitem [{\citenamefont {de~Sousa}\ and\ \citenamefont
  {Moore}(2008)}]{sousa:012406}%
  \BibitemOpen
  \bibfield  {author} {\bibinfo {author} {\bibfnamefont {R.}~\bibnamefont
  {de~Sousa}}\ and\ \bibinfo {author} {\bibfnamefont {J.~E.}\ \bibnamefont
  {Moore}},\ }\href@noop {} {\bibfield  {journal} {\bibinfo  {journal} {Phys.
  Rev. B}\ }\textbf {\bibinfo {volume} {77}},\ \bibinfo {pages} {012406}
  (\bibinfo {year} {2008})}\BibitemShut {NoStop}%
\bibitem [{\citenamefont {Cazayous}\ \emph {et~al.}(2008)\citenamefont
  {Cazayous}, \citenamefont {Gallais}, \citenamefont {Sacuto}, \citenamefont
  {de~Sousa}, \citenamefont {Lebeugle},\ and\ \citenamefont
  {Colson}}]{Cazayous}%
  \BibitemOpen
  \bibfield  {author} {\bibinfo {author} {\bibfnamefont {M.}~\bibnamefont
  {Cazayous}}, \bibinfo {author} {\bibfnamefont {Y.}~\bibnamefont {Gallais}},
  \bibinfo {author} {\bibfnamefont {A.}~\bibnamefont {Sacuto}}, \bibinfo
  {author} {\bibfnamefont {R.}~\bibnamefont {de~Sousa}}, \bibinfo {author}
  {\bibfnamefont {D.}~\bibnamefont {Lebeugle}}, \ and\ \bibinfo {author}
  {\bibfnamefont {D.}~\bibnamefont {Colson}},\ }\href@noop {} {\bibfield
  {journal} {\bibinfo  {journal} {Phys. Rev. Lett.}\ }\textbf {\bibinfo
  {volume} {101}},\ \bibinfo {pages} {037601} (\bibinfo {year}
  {2008})}\BibitemShut {NoStop}%
\bibitem [{\citenamefont {Kumar}\ \emph {et~al.}(2008)\citenamefont {Kumar},
  \citenamefont {Murari},\ and\ \citenamefont {Katiyara}}]{Kumar1}%
  \BibitemOpen
  \bibfield  {author} {\bibinfo {author} {\bibfnamefont {A.}~\bibnamefont
  {Kumar}}, \bibinfo {author} {\bibfnamefont {N.~M.}\ \bibnamefont {Murari}}, \
  and\ \bibinfo {author} {\bibfnamefont {R.~S.}\ \bibnamefont {Katiyara}},\
  }\href@noop {} {\bibfield  {journal} {\bibinfo  {journal} {Appl. Phys.
  Lett.}\ }\textbf {\bibinfo {volume} {92}},\ \bibinfo {pages} {152907}
  (\bibinfo {year} {2008})}\BibitemShut {NoStop}%
\bibitem [{\citenamefont {Rovillain}\ \emph {et~al.}(2010)\citenamefont
  {Rovillain}, \citenamefont {de~Sousa}, \citenamefont {Gallais}, \citenamefont
  {Sacuto}, \citenamefont {Measson}, \citenamefont {Colson}, \citenamefont
  {Forget}, \citenamefont {Bibes}, \citenamefont {Barthelemy},\ and\
  \citenamefont {Cazayous}}]{Rovillain}%
  \BibitemOpen
  \bibfield  {author} {\bibinfo {author} {\bibfnamefont {P.}~\bibnamefont
  {Rovillain}}, \bibinfo {author} {\bibfnamefont {R.}~\bibnamefont {de~Sousa}},
  \bibinfo {author} {\bibfnamefont {Y.}~\bibnamefont {Gallais}}, \bibinfo
  {author} {\bibfnamefont {A.}~\bibnamefont {Sacuto}}, \bibinfo {author}
  {\bibfnamefont {M.~A.}\ \bibnamefont {Measson}}, \bibinfo {author}
  {\bibfnamefont {D.}~\bibnamefont {Colson}}, \bibinfo {author} {\bibfnamefont
  {A.}~\bibnamefont {Forget}}, \bibinfo {author} {\bibfnamefont
  {M.}~\bibnamefont {Bibes}}, \bibinfo {author} {\bibfnamefont
  {A.}~\bibnamefont {Barthelemy}}, \ and\ \bibinfo {author} {\bibfnamefont
  {M.}~\bibnamefont {Cazayous}},\ }\href@noop {} {\bibfield  {journal}
  {\bibinfo  {journal} {Nature Materials}\ }\textbf {\bibinfo {volume} {9}},\
  \bibinfo {pages} {975} (\bibinfo {year} {2010})}\BibitemShut {NoStop}%
\bibitem [{\citenamefont {Kumar}\ \emph {et~al.}(2011)\citenamefont {Kumar},
  \citenamefont {Scott},\ and\ \citenamefont {Katiyar}}]{BiFeO3_film_field}%
  \BibitemOpen
  \bibfield  {author} {\bibinfo {author} {\bibfnamefont {A.}~\bibnamefont
  {Kumar}}, \bibinfo {author} {\bibfnamefont {J.~F.}\ \bibnamefont {Scott}}, \
  and\ \bibinfo {author} {\bibfnamefont {R.~S.}\ \bibnamefont {Katiyar}},\
  }\href@noop {} {\bibfield  {journal} {\bibinfo  {journal} {Appl. Phys.
  Lett.}\ }\textbf {\bibinfo {volume} {99}},\ \bibinfo {pages} {062504}
  (\bibinfo {year} {2011})}\BibitemShut {NoStop}%
\bibitem [{\citenamefont {Hlinka}\ \emph {et~al.}(2011)\citenamefont {Hlinka},
  \citenamefont {Pokorny}, \citenamefont {Karimi},\ and\ \citenamefont
  {Reaney}}]{BFO_Raman1}%
  \BibitemOpen
  \bibfield  {author} {\bibinfo {author} {\bibfnamefont {J.}~\bibnamefont
  {Hlinka}}, \bibinfo {author} {\bibfnamefont {J.}~\bibnamefont {Pokorny}},
  \bibinfo {author} {\bibfnamefont {S.}~\bibnamefont {Karimi}}, \ and\ \bibinfo
  {author} {\bibfnamefont {I.~M.}\ \bibnamefont {Reaney}},\ }\href@noop {}
  {\bibfield  {journal} {\bibinfo  {journal} {Phys. Rev. B}\ }\textbf {\bibinfo
  {volume} {83}},\ \bibinfo {pages} {020101} (\bibinfo {year}
  {2011})}\BibitemShut {NoStop}%
\bibitem [{\citenamefont {Palai}\ \emph {et~al.}(2010)\citenamefont {Palai},
  \citenamefont {Schmid}, \citenamefont {Scott},\ and\ \citenamefont
  {Katiyar}}]{BFO_Raman2}%
  \BibitemOpen
  \bibfield  {author} {\bibinfo {author} {\bibfnamefont {R.}~\bibnamefont
  {Palai}}, \bibinfo {author} {\bibfnamefont {H.}~\bibnamefont {Schmid}},
  \bibinfo {author} {\bibfnamefont {J.~F.}\ \bibnamefont {Scott}}, \ and\
  \bibinfo {author} {\bibfnamefont {R.~S.}\ \bibnamefont {Katiyar}},\
  }\href@noop {} {\bibfield  {journal} {\bibinfo  {journal} {Phys. Rev. B}\
  }\textbf {\bibinfo {volume} {81}},\ \bibinfo {pages} {064110} (\bibinfo
  {year} {2010})}\BibitemShut {NoStop}%
\bibitem [{\citenamefont {Haumont}\ \emph {et~al.}(2006)\citenamefont
  {Haumont}, \citenamefont {Kreisel}, \citenamefont {Bouvier},\ and\
  \citenamefont {Hippert}}]{haumont:132101}%
  \BibitemOpen
  \bibfield  {author} {\bibinfo {author} {\bibfnamefont {R.}~\bibnamefont
  {Haumont}}, \bibinfo {author} {\bibfnamefont {J.}~\bibnamefont {Kreisel}},
  \bibinfo {author} {\bibfnamefont {P.}~\bibnamefont {Bouvier}}, \ and\
  \bibinfo {author} {\bibfnamefont {F.}~\bibnamefont {Hippert}},\ }\href@noop
  {} {\bibfield  {journal} {\bibinfo  {journal} {Phys. Rev. B}\ }\textbf
  {\bibinfo {volume} {73}},\ \bibinfo {pages} {132101} (\bibinfo {year}
  {2006})}\BibitemShut {NoStop}%
\bibitem [{\citenamefont {Rovillain}\ \emph {et~al.}(2009)\citenamefont
  {Rovillain}, \citenamefont {Cazayous}, \citenamefont {Gallais}, \citenamefont
  {Sacuto}, \citenamefont {Lobo}, \citenamefont {Lebeugle},\ and\ \citenamefont
  {Colson}}]{Rovillain_Phonon}%
  \BibitemOpen
  \bibfield  {author} {\bibinfo {author} {\bibfnamefont {P.}~\bibnamefont
  {Rovillain}}, \bibinfo {author} {\bibfnamefont {M.}~\bibnamefont {Cazayous}},
  \bibinfo {author} {\bibfnamefont {Y.}~\bibnamefont {Gallais}}, \bibinfo
  {author} {\bibfnamefont {A.}~\bibnamefont {Sacuto}}, \bibinfo {author}
  {\bibfnamefont {R.}~\bibnamefont {Lobo}}, \bibinfo {author} {\bibfnamefont
  {D.}~\bibnamefont {Lebeugle}}, \ and\ \bibinfo {author} {\bibfnamefont
  {D.}~\bibnamefont {Colson}},\ }\href@noop {} {\bibfield  {journal} {\bibinfo
  {journal} {Phys. Rev. B}\ }\textbf {\bibinfo {volume} {79}},\ \bibinfo
  {pages} {180411} (\bibinfo {year} {2009})}\BibitemShut {NoStop}%
\bibitem [{\citenamefont {Lobo}\ \emph {et~al.}(2007)\citenamefont {Lobo},
  \citenamefont {Moreira}, \citenamefont {Lebeugle},\ and\ \citenamefont
  {Colson}}]{Lobo_infrared}%
  \BibitemOpen
  \bibfield  {author} {\bibinfo {author} {\bibfnamefont {R.~P. S.~M.}\
  \bibnamefont {Lobo}}, \bibinfo {author} {\bibfnamefont {R.~L.}\ \bibnamefont
  {Moreira}}, \bibinfo {author} {\bibfnamefont {D.}~\bibnamefont {Lebeugle}}, \
  and\ \bibinfo {author} {\bibfnamefont {D.}~\bibnamefont {Colson}},\
  }\href@noop {} {\bibfield  {journal} {\bibinfo  {journal} {Phys. Rev. B}\
  }\textbf {\bibinfo {volume} {76}},\ \bibinfo {pages} {172105} (\bibinfo
  {year} {2007})}\BibitemShut {NoStop}%
\bibitem [{\citenamefont {Xu}\ \emph {et~al.}(2012)\citenamefont {Xu},
  \citenamefont {Wen}, \citenamefont {Berlijn}, \citenamefont {Gehring},
  \citenamefont {Stock}, \citenamefont {Stone}, \citenamefont {Ku},
  \citenamefont {Gu}, \citenamefont {Shapiro}, \citenamefont {Birgeneau},\ and\
  \citenamefont {Xu}}]{BFO_Xu}%
  \BibitemOpen
  \bibfield  {author} {\bibinfo {author} {\bibfnamefont {Z.}~\bibnamefont
  {Xu}}, \bibinfo {author} {\bibfnamefont {J.}~\bibnamefont {Wen}}, \bibinfo
  {author} {\bibfnamefont {T.}~\bibnamefont {Berlijn}}, \bibinfo {author}
  {\bibfnamefont {P.~M.}\ \bibnamefont {Gehring}}, \bibinfo {author}
  {\bibfnamefont {C.}~\bibnamefont {Stock}}, \bibinfo {author} {\bibfnamefont
  {M.~B.}\ \bibnamefont {Stone}}, \bibinfo {author} {\bibfnamefont
  {W.}~\bibnamefont {Ku}}, \bibinfo {author} {\bibfnamefont {G.}~\bibnamefont
  {Gu}}, \bibinfo {author} {\bibfnamefont {S.~M.}\ \bibnamefont {Shapiro}},
  \bibinfo {author} {\bibfnamefont {R.~J.}\ \bibnamefont {Birgeneau}}, \ and\
  \bibinfo {author} {\bibfnamefont {G.}~\bibnamefont {Xu}},\ }\href@noop {}
  {\bibfield  {journal} {\bibinfo  {journal} {Phys. Rev. B}\ }\textbf {\bibinfo
  {volume} {86}},\ \bibinfo {pages} {174419} (\bibinfo {year}
  {2012})}\BibitemShut {NoStop}%
\bibitem [{\citenamefont {Jeong}\ \emph {et~al.}(2012)\citenamefont {Jeong},
  \citenamefont {Goremychkin}, \citenamefont {Guidi}, \citenamefont {Nakajima},
  \citenamefont {Jeon}, \citenamefont {Kim}, \citenamefont {Furukawa},
  \citenamefont {Kim}, \citenamefont {Lee}, \citenamefont {Kiryukhin},
  \citenamefont {Cheong},\ and\ \citenamefont {Park}}]{cheong}%
  \BibitemOpen
  \bibfield  {author} {\bibinfo {author} {\bibfnamefont {J.}~\bibnamefont
  {Jeong}}, \bibinfo {author} {\bibfnamefont {E.~A.}\ \bibnamefont
  {Goremychkin}}, \bibinfo {author} {\bibfnamefont {T.}~\bibnamefont {Guidi}},
  \bibinfo {author} {\bibfnamefont {K.}~\bibnamefont {Nakajima}}, \bibinfo
  {author} {\bibfnamefont {G.~S.}\ \bibnamefont {Jeon}}, \bibinfo {author}
  {\bibfnamefont {S.-A.}\ \bibnamefont {Kim}}, \bibinfo {author} {\bibfnamefont
  {S.}~\bibnamefont {Furukawa}}, \bibinfo {author} {\bibfnamefont {Y.~B.}\
  \bibnamefont {Kim}}, \bibinfo {author} {\bibfnamefont {S.}~\bibnamefont
  {Lee}}, \bibinfo {author} {\bibfnamefont {V.}~\bibnamefont {Kiryukhin}},
  \bibinfo {author} {\bibfnamefont {S.-W.}\ \bibnamefont {Cheong}}, \ and\
  \bibinfo {author} {\bibfnamefont {J.-G.}\ \bibnamefont {Park}},\ }\href
  {\doibase 10.1103/PhysRevLett.108.077202} {\bibfield  {journal} {\bibinfo
  {journal} {Phys. Rev. Lett.}\ }\textbf {\bibinfo {volume} {108}},\ \bibinfo
  {pages} {077202} (\bibinfo {year} {2012})}\BibitemShut {NoStop}%
\bibitem [{\citenamefont {Matsuda}\ \emph {et~al.}(2012)\citenamefont
  {Matsuda}, \citenamefont {Fishman}, \citenamefont {Hong}, \citenamefont
  {Lee}, \citenamefont {Ushiyama}, \citenamefont {Yangisawa}, \citenamefont
  {Tomioka},\ and\ \citenamefont {Ito}}]{Matsuda}%
  \BibitemOpen
  \bibfield  {author} {\bibinfo {author} {\bibfnamefont {M.}~\bibnamefont
  {Matsuda}}, \bibinfo {author} {\bibfnamefont {R.~S.}\ \bibnamefont
  {Fishman}}, \bibinfo {author} {\bibfnamefont {T.}~\bibnamefont {Hong}},
  \bibinfo {author} {\bibfnamefont {C.~H.}\ \bibnamefont {Lee}}, \bibinfo
  {author} {\bibfnamefont {T.}~\bibnamefont {Ushiyama}}, \bibinfo {author}
  {\bibfnamefont {Y.}~\bibnamefont {Yangisawa}}, \bibinfo {author}
  {\bibfnamefont {Y.}~\bibnamefont {Tomioka}}, \ and\ \bibinfo {author}
  {\bibfnamefont {T.}~\bibnamefont {Ito}},\ }\href@noop {} {\bibfield
  {journal} {\bibinfo  {journal} {Phys. Rev. Lett.}\ }\textbf {\bibinfo
  {volume} {109}},\ \bibinfo {pages} {067205} (\bibinfo {year}
  {2012})}\BibitemShut {NoStop}%
\bibitem [{\citenamefont {Delaire}\ \emph {et~al.}(2012)\citenamefont
  {Delaire}, \citenamefont {Stone}, \citenamefont {Ma}, \citenamefont {Huq},
  \citenamefont {Gout}, \citenamefont {Brown}, \citenamefont {Wang},\ and\
  \citenamefont {Ren}}]{Delaire}%
  \BibitemOpen
  \bibfield  {author} {\bibinfo {author} {\bibfnamefont {O.}~\bibnamefont
  {Delaire}}, \bibinfo {author} {\bibfnamefont {M.~B.}\ \bibnamefont {Stone}},
  \bibinfo {author} {\bibfnamefont {J.}~\bibnamefont {Ma}}, \bibinfo {author}
  {\bibfnamefont {A.}~\bibnamefont {Huq}}, \bibinfo {author} {\bibfnamefont
  {D.}~\bibnamefont {Gout}}, \bibinfo {author} {\bibfnamefont {C.}~\bibnamefont
  {Brown}}, \bibinfo {author} {\bibfnamefont {K.~F.}\ \bibnamefont {Wang}}, \
  and\ \bibinfo {author} {\bibfnamefont {Z.~F.}\ \bibnamefont {Ren}},\ }\href
  {\doibase 10.1103/PhysRevB.85.064405} {\bibfield  {journal} {\bibinfo
  {journal} {Phys. Rev. B}\ }\textbf {\bibinfo {volume} {85}},\ \bibinfo
  {pages} {064405} (\bibinfo {year} {2012})}\BibitemShut {NoStop}%
\bibitem [{\citenamefont {Borissenko}\ \emph {et~al.}(2013)\citenamefont
  {Borissenko}, \citenamefont {Goffinet}, \citenamefont {Bosak}, \citenamefont
  {Rovillain}, \citenamefont {Cazayous}, \citenamefont {Colson}, \citenamefont
  {Ghosez},\ and\ \citenamefont {Krisch}}]{BFO_IXS}%
  \BibitemOpen
  \bibfield  {author} {\bibinfo {author} {\bibfnamefont {E.}~\bibnamefont
  {Borissenko}}, \bibinfo {author} {\bibfnamefont {M.}~\bibnamefont
  {Goffinet}}, \bibinfo {author} {\bibfnamefont {A.}~\bibnamefont {Bosak}},
  \bibinfo {author} {\bibfnamefont {P.}~\bibnamefont {Rovillain}}, \bibinfo
  {author} {\bibfnamefont {M.}~\bibnamefont {Cazayous}}, \bibinfo {author}
  {\bibfnamefont {D.}~\bibnamefont {Colson}}, \bibinfo {author} {\bibfnamefont
  {P.}~\bibnamefont {Ghosez}}, \ and\ \bibinfo {author} {\bibfnamefont
  {M.}~\bibnamefont {Krisch}},\ }\href@noop {} {\bibfield  {journal} {\bibinfo
  {journal} {J. Phys. Cond. Matt.}\ }\textbf {\bibinfo {volume} {25}},\
  \bibinfo {pages} {102201} (\bibinfo {year} {2013})}\BibitemShut {NoStop}%
\bibitem [{\citenamefont {Ito}\ \emph {et~al.}(2011)\citenamefont {Ito},
  \citenamefont {Ushiyama}, \citenamefont {Yanagisawa}, \citenamefont {Kumai},\
  and\ \citenamefont {Tomioka}}]{Ito_BFO}%
  \BibitemOpen
  \bibfield  {author} {\bibinfo {author} {\bibfnamefont {T.}~\bibnamefont
  {Ito}}, \bibinfo {author} {\bibfnamefont {T.}~\bibnamefont {Ushiyama}},
  \bibinfo {author} {\bibfnamefont {Y.}~\bibnamefont {Yanagisawa}}, \bibinfo
  {author} {\bibfnamefont {R.}~\bibnamefont {Kumai}}, \ and\ \bibinfo {author}
  {\bibfnamefont {Y.}~\bibnamefont {Tomioka}},\ }\href@noop {} {\bibfield
  {journal} {\bibinfo  {journal} {Crystal Growth \& Design}\ }\textbf {\bibinfo
  {volume} {11}},\ \bibinfo {pages} {5139} (\bibinfo {year}
  {2011})}\BibitemShut {NoStop}%
\bibitem [{\citenamefont {Stone}\ \emph {et~al.}(2014)\citenamefont {Stone},
  \citenamefont {Niedziela}, \citenamefont {Abernathy}, \citenamefont
  {DeBeer-Schmitt}, \citenamefont {Ehlers}, \citenamefont {Garlea},
  \citenamefont {Granroth}, \citenamefont {Graves-Brook}, \citenamefont
  {Kolesnikov}, \citenamefont {Podlesnyak},\ and\ \citenamefont
  {Winn}}]{HYSPEC}%
  \BibitemOpen
  \bibfield  {author} {\bibinfo {author} {\bibfnamefont {M.~B.}\ \bibnamefont
  {Stone}}, \bibinfo {author} {\bibfnamefont {J.~L.}\ \bibnamefont
  {Niedziela}}, \bibinfo {author} {\bibfnamefont {D.~L.}\ \bibnamefont
  {Abernathy}}, \bibinfo {author} {\bibfnamefont {L.}~\bibnamefont
  {DeBeer-Schmitt}}, \bibinfo {author} {\bibfnamefont {G.}~\bibnamefont
  {Ehlers}}, \bibinfo {author} {\bibfnamefont {O.}~\bibnamefont {Garlea}},
  \bibinfo {author} {\bibfnamefont {G.~E.}\ \bibnamefont {Granroth}}, \bibinfo
  {author} {\bibfnamefont {M.}~\bibnamefont {Graves-Brook}}, \bibinfo {author}
  {\bibfnamefont {A.~I.}\ \bibnamefont {Kolesnikov}}, \bibinfo {author}
  {\bibfnamefont {A.}~\bibnamefont {Podlesnyak}}, \ and\ \bibinfo {author}
  {\bibfnamefont {B.}~\bibnamefont {Winn}},\ }\href@noop {} {\bibfield
  {journal} {\bibinfo  {journal} {Rev. of Sci. Inst.}\ }\textbf {\bibinfo
  {volume} {85}},\ \bibinfo {pages} {045113} (\bibinfo {year}
  {2014})}\BibitemShut {NoStop}%
\bibitem [{\citenamefont {Neighbours}\ and\ \citenamefont
  {Schacher}(1967)}]{ElasticConstant}%
  \BibitemOpen
  \bibfield  {author} {\bibinfo {author} {\bibfnamefont {J.~R.}\ \bibnamefont
  {Neighbours}}\ and\ \bibinfo {author} {\bibfnamefont {G.~E.}\ \bibnamefont
  {Schacher}},\ }\href@noop {} {\bibfield  {journal} {\bibinfo  {journal} {J.
  Appl. Phys.}\ }\textbf {\bibinfo {volume} {38}},\ \bibinfo {pages} {5366}
  (\bibinfo {year} {1967})}\BibitemShut {NoStop}%
\bibitem [{\citenamefont {Smirnova}\ \emph {et~al.}(2011)\citenamefont
  {Smirnova}, \citenamefont {Sotnikov}, \citenamefont {Ktitorov}, \citenamefont
  {Zaitseva}, \citenamefont {Schmidt},\ and\ \citenamefont
  {Weihnacht}}]{Smirnova_Acoustic}%
  \BibitemOpen
  \bibfield  {author} {\bibinfo {author} {\bibfnamefont {E.~P.}\ \bibnamefont
  {Smirnova}}, \bibinfo {author} {\bibfnamefont {A.}~\bibnamefont {Sotnikov}},
  \bibinfo {author} {\bibfnamefont {S.}~\bibnamefont {Ktitorov}}, \bibinfo
  {author} {\bibfnamefont {N.}~\bibnamefont {Zaitseva}}, \bibinfo {author}
  {\bibfnamefont {H.}~\bibnamefont {Schmidt}}, \ and\ \bibinfo {author}
  {\bibfnamefont {M.}~\bibnamefont {Weihnacht}},\ }\href@noop {} {\bibfield
  {journal} {\bibinfo  {journal} {Euro. Phys. J. B}\ }\textbf {\bibinfo
  {volume} {83}},\ \bibinfo {pages} {39} (\bibinfo {year} {2011})}\BibitemShut
  {NoStop}%
\bibitem [{\citenamefont {Doig}\ \emph {et~al.}(2013)\citenamefont {Doig},
  \citenamefont {Aguesse}, \citenamefont {Axelsson}, \citenamefont {Alford},
  \citenamefont {Nawaz}, \citenamefont {Palkar}, \citenamefont {Jones},
  \citenamefont {Johnson}, \citenamefont {Synowicki},\ and\ \citenamefont
  {Lloyd-Hughes}}]{DoigK_phonon}%
  \BibitemOpen
  \bibfield  {author} {\bibinfo {author} {\bibfnamefont {K.~I.}\ \bibnamefont
  {Doig}}, \bibinfo {author} {\bibfnamefont {F.}~\bibnamefont {Aguesse}},
  \bibinfo {author} {\bibfnamefont {A.~K.}\ \bibnamefont {Axelsson}}, \bibinfo
  {author} {\bibfnamefont {N.~M.}\ \bibnamefont {Alford}}, \bibinfo {author}
  {\bibfnamefont {S.}~\bibnamefont {Nawaz}}, \bibinfo {author} {\bibfnamefont
  {V.~R.}\ \bibnamefont {Palkar}}, \bibinfo {author} {\bibfnamefont {S.~P.~P.}\
  \bibnamefont {Jones}}, \bibinfo {author} {\bibfnamefont {R.~D.}\ \bibnamefont
  {Johnson}}, \bibinfo {author} {\bibfnamefont {R.~A.}\ \bibnamefont
  {Synowicki}}, \ and\ \bibinfo {author} {\bibfnamefont {J.}~\bibnamefont
  {Lloyd-Hughes}},\ }\href@noop {} {\bibfield  {journal} {\bibinfo  {journal}
  {Phys. Rev. B}\ }\textbf {\bibinfo {volume} {88}},\ \bibinfo {pages} {094425}
  (\bibinfo {year} {2013})}\BibitemShut {NoStop}%
\bibitem [{\citenamefont {Goffinet}\ \emph {et~al.}(2009)\citenamefont
  {Goffinet}, \citenamefont {Hermet}, \citenamefont {Bilc},\ and\ \citenamefont
  {Ghosez}}]{Goffinet}%
  \BibitemOpen
  \bibfield  {author} {\bibinfo {author} {\bibfnamefont {M.}~\bibnamefont
  {Goffinet}}, \bibinfo {author} {\bibfnamefont {P.}~\bibnamefont {Hermet}},
  \bibinfo {author} {\bibfnamefont {D.~I.}\ \bibnamefont {Bilc}}, \ and\
  \bibinfo {author} {\bibfnamefont {P.}~\bibnamefont {Ghosez}},\ }\href@noop {}
  {\bibfield  {journal} {\bibinfo  {journal} {Phys. Rev. B}\ }\textbf {\bibinfo
  {volume} {79}},\ \bibinfo {pages} {014403} (\bibinfo {year}
  {2009})}\BibitemShut {NoStop}%
\bibitem [{\citenamefont {Shang}\ \emph {et~al.}(2009)\citenamefont {Shang},
  \citenamefont {Sheng}, \citenamefont {Wang}, \citenamefont {Chen},\ and\
  \citenamefont {Liu}}]{Shang_BFOelastic}%
  \BibitemOpen
  \bibfield  {author} {\bibinfo {author} {\bibfnamefont {S.~L.}\ \bibnamefont
  {Shang}}, \bibinfo {author} {\bibfnamefont {G.}~\bibnamefont {Sheng}},
  \bibinfo {author} {\bibfnamefont {Y.}~\bibnamefont {Wang}}, \bibinfo {author}
  {\bibfnamefont {L.~Q.}\ \bibnamefont {Chen}}, \ and\ \bibinfo {author}
  {\bibfnamefont {Z.~K.}\ \bibnamefont {Liu}},\ }\href@noop {} {\bibfield
  {journal} {\bibinfo  {journal} {Phys. Rev. B}\ }\textbf {\bibinfo {volume}
  {80}},\ \bibinfo {pages} {052102} (\bibinfo {year} {2009})}\BibitemShut
  {NoStop}%
\bibitem [{\citenamefont {Stock}\ \emph {et~al.}(2013)\citenamefont {Stock},
  \citenamefont {Dunsiger}, \citenamefont {Mole}, \citenamefont {Li},\ and\
  \citenamefont {Luo}}]{Stock_PFN}%
  \BibitemOpen
  \bibfield  {author} {\bibinfo {author} {\bibfnamefont {C.}~\bibnamefont
  {Stock}}, \bibinfo {author} {\bibfnamefont {S.~R.}\ \bibnamefont {Dunsiger}},
  \bibinfo {author} {\bibfnamefont {R.~A.}\ \bibnamefont {Mole}}, \bibinfo
  {author} {\bibfnamefont {X.}~\bibnamefont {Li}}, \ and\ \bibinfo {author}
  {\bibfnamefont {H.}~\bibnamefont {Luo}},\ }\href {\doibase
  10.1103/PhysRevB.88.094105} {\bibfield  {journal} {\bibinfo  {journal} {Phys.
  Rev. B}\ }\textbf {\bibinfo {volume} {88}},\ \bibinfo {pages} {094105}
  (\bibinfo {year} {2013})}\BibitemShut {NoStop}%
\end{thebibliography}

%

\end{document}